\documentclass[a4paper,11pt]{article}
\usepackage{jheppub} % for details on the use of the package, please see the JINST-author-manual

\arxivnumber{2510.19916}

\usepackage[T1]{fontenc}
\usepackage[dvipsnames]{xcolor}
\usepackage{orcidlink}
\usepackage[english]{babel}
\usepackage{graphicx}
\usepackage{amsmath,amssymb}
\definecolor{myred}{RGB}{179, 27, 27}
\definecolor{teal}{RGB}{0, 128, 128}
\usepackage{hyperref}
\usepackage{rotating}
\hypersetup{
    colorlinks=true,
    linkcolor=teal, 
    citecolor=teal, 
    urlcolor=teal  
 }
\title{\boldmath Constraining the Swift Memory Burden Effect with GW250114-like Events}

\author{Chen Yuan\orcidlink{0000-0001-8560-5487}}
\affiliation{CENTRA, Departamento de Física, Instituto Superior Técnico – IST, Universidade de Lisboa – UL, Avenida Rovisco Pais 1, 1049–001 Lisboa, Portugal}

\author{Richard Brito\orcidlink{0000-0002-7807-3053}}
\affiliation{CENTRA, Departamento de Física, Instituto Superior Técnico – IST, Universidade de Lisboa – UL, Avenida Rovisco Pais 1, 1049–001 Lisboa, Portugal}

\abstract{
Black hole spectroscopy allows to infer the properties of the remnant of a binary black hole coalescence. Motivated by the recent proposal that a black hole's information load can alter its classical response to small perturbations, an effect known as the swift memory burden (SMB), we develop a minimal phenomenological framework to analyze the ringdown of a binary black hole merger and confront it with the data from the GW250114 event. In particular, we model the SMB imprint through a suppression factor derived as an upper bound and adopt the least-suppressive form. We perform a Bayesian analysis combining the frequencies of the (220) and (440) quasi-normal modes and obtain a lower bound $\log_{10}p \gtrsim 2$, where $p$ controls how the gaps reopen when the black hole's master mode occupation departs from the critical value. Moreover, using a Fisher information matrix (high signal-to-noise ratio) approximation, we forecast the lower bound $\log_{10}p \gtrsim 3$ for a GW250114-like event observed with Cosmic Explorer or Einstein Telescope. Within this minimal, weak-effect framework, our results disfavour rapid gap reopening, shedding light on how the swift memory burden effect can be probed with current and next-generation detectors.
}

\begin{document}

\maketitle
\flushbottom

\section{Introduction}

The detection of more than two hundred gravitational wave (GW) events \cite{ LIGOScientific:2018mvr, LIGOScientific:2020ibl, KAGRA:2021vkt,LIGOScientific:2025slb} has opened a new window to explore fundamental physics and the Universe. The growing number of GW events has enabled various stringent tests of Einstein's general relativity (GR) in the strong field regime~\cite{LIGOScientific:2016lio,LIGOScientific:2018dkp,LIGOScientific:2019fpa,LIGOScientific:2020tif,LIGOScientific:2021sio,KAGRA:2025oiz,LIGOScientific:2025obp}. Among all the GW events reported by the LIGO–
Virgo–KAGRA (LVK) collaboration so far, the recent event GW250114 was reported to be the loudest GW detection~\cite{KAGRA:2025oiz,LIGOScientific:2025obp}. Its exceptional network signal-to-noise ratio enables high-precision ringdown studies and stringent constraints on quasi-normal mode (QNM) spectra~\cite{KAGRA:2025oiz,LIGOScientific:2025obp}. For GW250114, at least three QNMs are identified or constrained, the quadrupolar 220 fundamental and first overtone 221, and
the hexadecapolar 440 mode, with the later being identified in an analysis that uses a parameterized inspiral–
merger–ringdown waveform~\cite{Brito:2018rfr,Ghosh:2021mrv,Maggio:2022hre,Pompili:2025cdc}. This unique event provides the sensitivity needed to probe potential departures from the Kerr ringdown pattern, such as small fractional shifts in the QNM frequencies that could encode departures from the Kerr black hole geometry or new physics operating in the near-horizon region \cite{Cardoso:2019rvt,Berti:2025hly}.

The ringdown phase of the signal emitted by a binary BH coalescence can be described by a linear superposition of exponentially damped sinusoids corresponding to QNMs of the final BH remnant. While closer to the merger, dynamical effects and nonlinearities might be important, they are typically subdominant to linear QNMs shortly after the merger~\cite{London:2014cma,Cheung:2022rbm,Mitman:2022qdl,Zhu:2024dyl,Chavda:2024awq,Ma:2024qcv,DeAmicis:2025xuh}. Within classical GR, the QNM frequencies and damping times only depend on the mass and spin of the BH remnant~\cite{Berti:2009kk} and therefore the measurement of these frequencies and damping times provides a powerful method to test the nature of the final object~\cite{Berti:2005ys,Dreyer:2003bv,Cardoso:2019mqo,Berti:2025hly}. The main modifications from the Kerr BH QNM spectra due to e.g. beyond GR effects, are expected to be captured by small shifts in the frequency and damping times with respect to the QNMs in GR, which have been computed in different beyond GR theories, see e.g.~\cite{Cano:2024jkd,Cano:2024ezp,Khoo:2024agm,Blazquez-Salcedo:2024oek,Chung:2025gyg,Chung:2025wbg} for some recent results. The approach of considering shifts in the frequencies and damping times to measure departures from the Kerr QNM spectra has been extensively used in GW data analysis~\cite{Gossan:2011ha,Meidam:2014jpa,Carullo:2018sfu,Brito:2018rfr,LIGOScientific:2020tif,LIGOScientific:2021sio,LIGOScientific:2025obp} since it provides a simple theory-agnostic way to test GR in the ringdown phase and can easily be used to constrain specific beyond GR theories, as done in~\cite{Chung:2025wbg} (see however~\cite{Crescimbeni:2024sam,Lestingi:2025jyb,Crescimbeni:2025kxi} for a proposal of a more general parametrization for a beyond GR ringdown model that includes modes induced by extra fields).

The first GW event detected by the LIGO detectors, GW150914~\cite{LIGOScientific:2016aoc}, provided the first example of the measurement of the frequency and damping time of the least damped QNM (the 220 mode) of a merger remnant~\cite{LIGOScientific:2016lio} which was followed by claims that the post-peak GW signal from GW150914 contains evidence for the presence of the first quadrupolar overtone 221~\cite{Isi:2019aib} (see however Refs.~\cite{Finch:2022ynt,Cotesta:2022pci,Isi:2023nif,Carullo:2023gtf,Baibhav:2023clw} for discussions on the significance of this detection). Since then, similar analyses have been done by the LVK Collaboration on a larger number of GW events~\cite{LIGOScientific:2020tif,LIGOScientific:2021sio,LIGOScientific:2025obp}, with GW250114 providing the most stringent constraints on departures from the Kerr QNM spectra so far~\cite{LIGOScientific:2025obp}. Notably, using an inspiral-merger-ringdown (IMR) model that introduces fractional deviations to the frequency and decay time of the fundamental QNMs, Ref.~\cite{LIGOScientific:2025obp} reported constraints on deviations from 220 frequency at the $\sim 2\%$  level.

Such stringent constraints motivate the analysis of models that can strongly modify the ringdown frequencies. In particular, it has recently been proposed that a mechanism~--~the swift memory burden (SMB)~--~ linking a BH's information load to its classical response under perturbations could have observable effects in the ringdown of a binary BH merger~ \cite{Dvali:2025sog}. In this framework, systems with exceptionally efficient information storage (black holes being the typical example) feature nearly-gapless ``memory modes'' whose gaps are lowered by highly occupied, macroscopic master modes. While different information loads are degenerate in the unperturbed state, a classical disturbance (e.g., a merger) drives the system away from the gapless point and the stored information then back-reacts and resists this evolution, thereby altering the classical dynamics of the perturbed black hole \cite{Dvali:2017nis, Dvali:2017ktv, Dvali:2018vvx, Dvali:2018xoc, Dvali:2018tqi,Dvali:2018xpy, Dvali:2018ytn, Dvali:2020wft}.

The SMB response can be summarized as an information-load–induced suppression of depletion of the BH’s master mode. In the assisted–gaplessness picture, highly occupied soft master modes render the memory modes gapless, allowing the BH to store an extensive number of patterns at negligible energy cost. When a classical perturbation (e.g., a merger) attempts to drive the system away from this critical point, the stored pattern resists the change, through which the information load back-reacts and modifies the master mode’s effective frequency. Crucially, this modification pushes the effective resonance towards the infrared. 
On the other hand, as soon as the master mode starts to deplete into a radiation channel, the burden reacts by detuning the would-be resonance and closing the channel again. The net effect is that only a small fraction of quanta can be emitted in any given frequency bin, producing a suppressed spectrum relative to the unburdened case. 
The SMB effect makes the depletion most strongly suppressed at the unburdened resonance, with the suppression relaxing as the frequency is redshifted. Consequently, the peak of the spectrum shifts to lower frequencies and the amplitude of the spectrum is suppressed near the unburdened frequency.

Guided by the SMB picture, our analysis proceeds in four steps. First, we provide a brief review of the SMB in Section~\ref{sec:smb}. In Section~\ref{sec:shift}, we develop a minimal single-mode ringdown model that includes the SMB-induced suppression and the QNM frequency redshift due to the SMB effect. Then in Section~\ref{sec:bayesian}, we confront this parameterization by performing a Bayesian parameter estimation using the QNM data from the GW250114 event~\cite{LIGOScientific:2025obp}. Next, we assess the prospects for next-generation detectors by computing Fisher-matrix forecasts \cite{Cutler:1994ys} for a GW250114-like source observed by Cosmic Explorer (CE) \cite{Hall:2020dps, Dwyer:2014fpa} or Einstein Telescope (ET) \cite{Punturo:2010zz,ET:2019dnz} in Section~\ref{sec:forecast}. Finally, we discuss and summarize our results in Section~\ref{sec:sum}.

\section{Swift memory burden}\label{sec:smb}
In this section, we will give a brief review of the SMB effect based on Ref.~\cite{Dvali:2025sog}. BHs are systems of high efficiency of information storage. In the microscopic picture, a BH belongs to a universal class of systems that achieve ``assisted gaplessness'': a small set of soft, highly populated ``master'' modes lowers the effective gaps of a large family of ``memory'' modes \cite{Dvali:2017nis, Dvali:2017ktv, Dvali:2018vvx, Dvali:2018xoc, Dvali:2018tqi,Dvali:2018xpy, Dvali:2018ytn, Dvali:2020wft}. At the critical occupation of the master mode, the memory-mode gaps collapse to zero, so that an exponentially large space of memory patterns becomes degenerate. As a result, the memory modes can store information at negligible energy cost.

The information stored in the gapless memory space back-reacts on the dynamics when the system is changing (e.g., losing energy through Hawking radiation). A convenient macroscopic control parameter is $\mu$, 
% \begin{equation}
% \mu \equiv \frac{E_{\rm ms}}{p\,E_{\rm P}}=\frac{m_\alpha N}{p\,E_{\rm P}},
% \end{equation}
which compares the energy invested in the master mode to the vacuum energy cost of the same memory pattern. Smaller $\mu$ means more efficient storage and a stronger back-reaction, while $\mu\to \infty $ refers to empty information load. Earlier studies emphasized the MB phase that emerges gradually during Hawking evaporation \cite{Dvali:2018xpy, Dvali:2018ytn, Dvali:2020wft}. Such an effect allows ultralight primordial black holes to survive from evaporation and to explain the dark matter in the universe \cite{Zantedeschi:2024ram,Franciolini:2023osw,Thoss:2024hsr,Alexandre:2024nuo,Balaji:2024hpu,Haque:2024eyh,Jiang:2024aju,Kohri:2024qpd,Chianese:2024rsn,Barker:2024mpz,Borah:2024bcr,Loc:2024qbz,Athron:2024fcj,Bandyopadhyay:2025ast,Calabrese:2025sfh,Boccia:2025hpm,Chianese:2025wrk,Dondarini:2025ktz,Chaudhuri:2025asm,Maity:2025ffa}.

However, the timescale of the MB is set by the timescale of the perturbation and it is typically negligible for astrophysical BHs. On the other hand, a recently proposed swift memory burden (SMB) effect points out that the information load could immediately affect a BH under a classical perturbation (e.g., a merger of binary BHs) \cite{Dvali:2025sog}. As argued in~\cite{Dvali:2025sog}, for astrophysical BHs formed from ordinary collapse and mergers, one expects $\mu\ll 1$, which might make the SMB imprint potentially observable in precision ringdown spectroscopy.

% parametrized by an angle $\vartheta_\beta$ via
% \begin{equation}
% \frac{\Delta n_\beta}{N_1} \equiv \sin^2\!\vartheta_\beta ,
% \end{equation}
% with the distribution
% \[
% \tan(2\vartheta_\beta)=\frac{2\tilde m_\beta}{M(\vartheta_\beta)} ,
% \qquad
% M(\vartheta_\beta)\equiv m_1\!\left(1-\frac{1}{\mu_1}\sin^{2p-2}\!\vartheta_\beta\right)-\omega_\beta .
% \]
The key result is that the effective master frequency is shifted toward the infrared by the memory burden, so exact resonance becomes difficult to maintain once information is loaded.
As depletion proceeds, the burden swiftly affect the system. Even if a momentary resonance is achieved, time evolution quickly drives it off resonance, narrowing the resonance and reducing the radiated quanta. As a result, the intensity of the radiation gets suppressed by a factor
\begin{equation}\label{Sfactor}
S(f;\mu,p)=
\begin{cases}
\displaystyle \Bigl[\mu\Bigl(1-\frac{f}{f_R}\Bigr)\Bigr]^{\frac{1}{p-1}}, & f\le f_R,\\[6pt]
1, & f\ge f_R.
\end{cases}
\end{equation}
where $f_R$ represents the unburdened radiation frequency and the free parameter $p>1$ controls how the gaps reopen when the master-mode occupation departs from the critical value.
In Ref.~\cite{Dvali:2025sog}, the intensity suppression appears as an upper bound, rather than an exact multiplicative factor. However, we will demonstrate that taking the upper bound as the suppression result provides a conservative constraint on the model parameter $p$. If the actual SMB induced suppression is stronger, one would expect to obtain more stringent bounds. Moreover, although Eq.~(\ref{Sfactor}) is singular at $p=1$ we restrict our analysis to $p>1$ where the modification from the SMB effect is weak as we explain below.

Note that the suppression factor $S$ for astrophysical BHs $(\mu \ll 1)$ is vanishingly small over a broad band if $p \sim \mathcal{O}(1)$, in particular during inspiral/merger where $f \ll f_R$, namely $S(f) < \mu^{{1}/(p-1)}\quad (\textrm{for }f < f_R)$.
Since the GW energy spectrum scales as $dE/df \propto D_L^2 f^2 |\tilde h(f)|^2$, where $D_L$ denotes the luminosity distance to the source and $h(f)$ is the frequency-domain strain, the suppression factor implies a strain-amplitude scaling $h(f)\to \sqrt{S}  h(f)$, and hence (at fixed detector noise) an approximate SNR scaling $\rho\to \sqrt{S}\rho$. To reproduce the GW250114 signal at its GR-inferred luminosity distance $D_L\simeq 400~\mathrm{Mpc}$ would therefore require a much smaller distance $D_L\rightarrow D_L\sqrt{S}$. This indicates the $p \sim \mathcal{O}(1)$ case would only be compatible with an extremely close-by source, which is astrophysically disfavored. On the other hand, compensating the amplitude suppression by increasing the intrinsic mass scale is also implausible. Using the inspiral scaling $h\propto \mathcal{M}_c^{5/6}/D_L$ would require $\mathcal{M}_c\to \mathcal{M}_c\,S^{-3/5}$, which corresponds to an enormous mass increase and would shift the signal to much lower frequencies, outside the LVK band. These order-of-magnitude estimates indicate that the $p \sim \mathcal{O}(1)$ case is strongly disfavored on detectability grounds even when $S$ is treated as the least-suppressive upper bound. We therefore focus on the weak-effect regime $p\gg 1$, where $S$ can be $\mathcal{O}(1)$ and the leading observable imprint is small.

\section{Peak shift due to SMB}\label{sec:shift}
We begin from the single-mode ringdown spectrum in the absence of SMB such that
\begin{equation}
\mathcal{E}_0(f)\;=\;\frac{A}{(f-f_R)^2+\Delta f^2},
\qquad 
\Delta f\equiv\frac{1}{2\pi\tau},
\end{equation}
where $A$ is the amplitude and we include the SMB suppression factor so that the total energy spectrum becomes
\begin{equation}
\mathcal{E}(f)\;=\;\mathcal{E}_0(f)\,S(f;\mu,p)\,.
\end{equation}
As mentioned above, the suppression factor is around unity for $p\gg1$ if the BH information load is full ($\mu\ll1$) and for $p\sim \mathcal{O}(1)$, the suppression factor is extremely small and almost no radiation will be emitted. Hence, a physical choice corresponds to $p\gg 1$. In this case, the peak of the ringdown spectrum will be redshifted due to the SMB suppression factor such that 
\begin{equation}
\frac{d\mathcal{E}(f)}{df} \Big|_{f=f_\star}\;=\; 0
\quad\Longrightarrow\quad
f_\star=f_R-\frac{1}{2\pi\tau\sqrt{2p-3}}.
\end{equation}
One can also obtain the dimensionless frequency shift, namely $f_R\to f_R(1+\delta f)$ which takes the form
\begin{equation}
    \delta f= -\frac{1}{2\pi f_R \tau \sqrt{2p-3}}.
\end{equation}
For astrophysical BHs formed from stellar collapse or binary BH mergers, the SMB effect can therefore imprint potentially observable signatures in the BH ringdown. Notably, the peak shift is independent of $\mu$ (which only rescales the amplitude). Therefore, a Bayesian parameter estimation of $\delta f$ will only place a bound on $p$.

We stress that Eq.~(\ref{Sfactor}) specifies an upper bound on the frequency-domain intensity, but does not provide the complex QNM eigenfrequency or a time-domain evolution equation.  Multiplying a Lorentzian by $S(f;\mu,p)$ produces a non-Lorentzian and asymmetric line shape. Although one may define an ``effective'' width of the spectrum, different definitions (for example, using the curvature at the peak or using the full width at half maximum) give different $\delta\tau_{\ell m n}$. The present minimal SMB model therefore does not make a unique prediction for a damping-time shift. In particular, we neither predict nor impose $\delta\tau_{\ell m n}=0$. 
Although a more complete analysis where both $\delta f_{\ell m n}$ and $\delta\tau_{\ell m n}$ are used simultaneously to constrain $p$ (and possibly $\mu$) would be preferable, here we instead marginalize over $\delta\tau_{\ell m n}$, only using $\delta f_{\ell m n}$ in our analysis.

\section{Bayesian inference based on GW250114}
\label{sec:bayesian}

To constrain the parameters of the SMB model, we perform a Bayesian parameter estimation analysis using the publicly available posteriors of the GW250114 event~\cite{KAGRA:2025oiz,LIGOScientific:2025obp}. 
We include both the dominant $l=m=2, n=0$ (220) mode and the subdominant $l=m=4, n=0$ (440) mode, for which the LVK Collaboration has reported the frequency deviations $\delta f_{220}$ and $\delta f_{440}$ relative to the General Relativity (GR) predictions, obtained using the \texttt{pSEOBNR} analysis~\cite{Brito:2018rfr,Ghosh:2021mrv,Maggio:2022hre,Pompili:2025cdc}, which  introduces fractional deviations to the frequency and decay time of the fundamental QNMs in the ringdown description of a full inspiral-merger-ringdown waveform model~\cite{LIGOScientific:2025obp}.

\begin{figure*}[ht]
    \centering
    \includegraphics[width=0.48\textwidth]{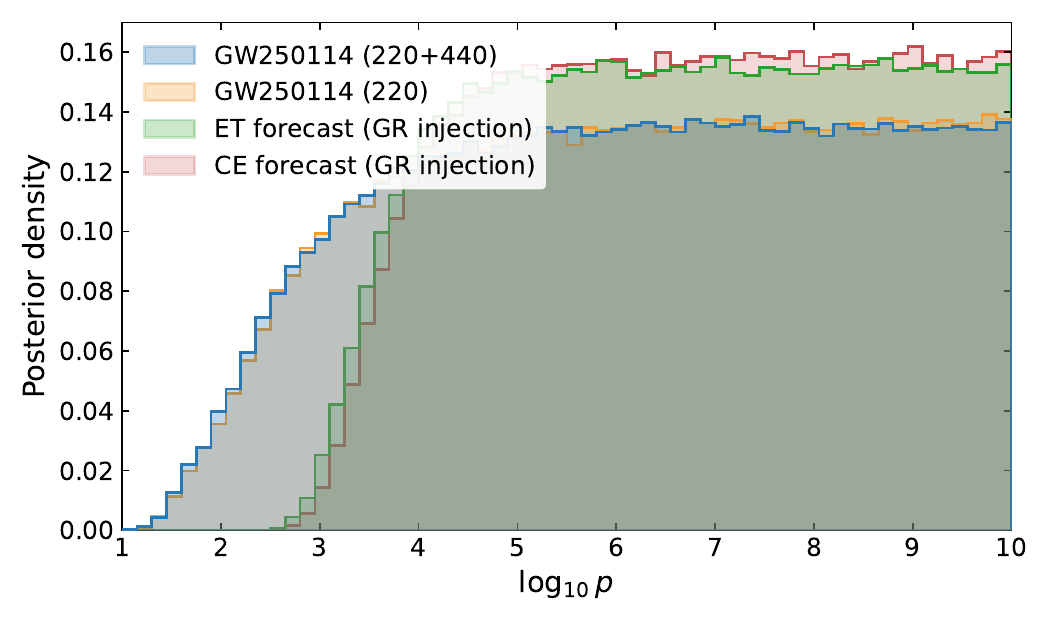}\hfill
    \includegraphics[width=0.48\textwidth]{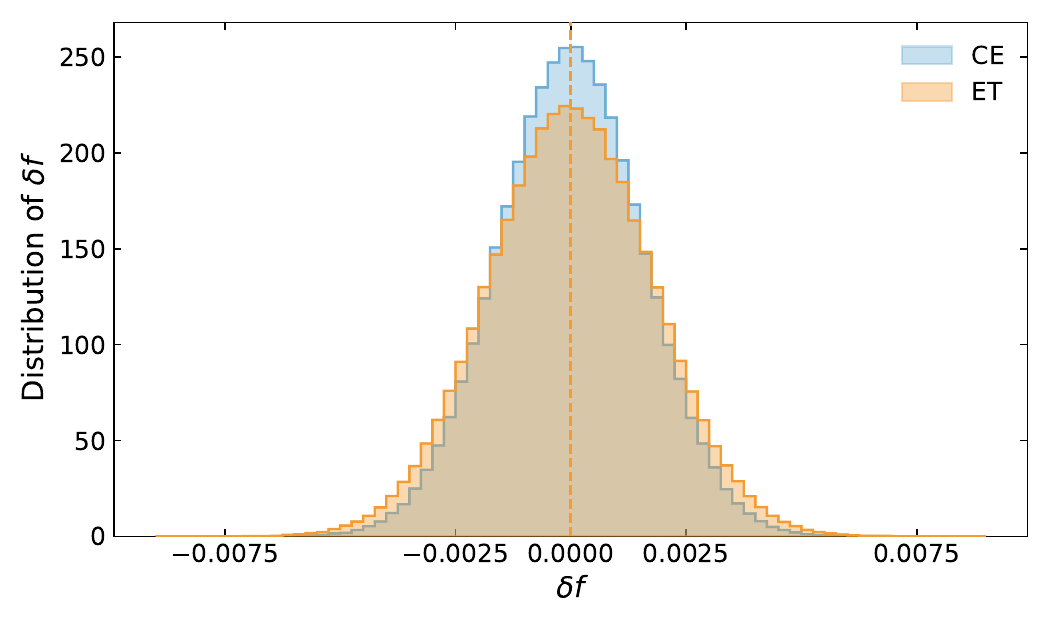}
    \caption{\label{fig:SMB_two_panel}
\textbf{Left panel:} Posterior histograms for $\log_{10}p$ from GW250114, using the 220 and 440 modes with a common-remnant marginalization (blue), and using only the 220 mode (orange).  The green and red histograms are Fisher forecasts for a GW250114-like source observed by ET and CE, respectively, with a GR injection $\delta f=0$.
\textbf{Right panel:} Marginalized histograms of the 220-mode fractional frequency deviation for the same ET and CE GR-injection forecasts.  The dashed lines mark the sample means, which coincide with the injected GR value $\delta f=0$.}
\end{figure*}

Following the LVK convention, we parametrize the fractional frequency shift as \cite{LIGOScientific:2025obp}
\begin{equation}
f_{\ell m n} \;=\; f^{\mathrm{GR}}_{\ell m n}\,\bigl( 1 + \delta_{\ell m n} (\boldsymbol{\theta}) \bigr),
\qquad
\delta_{\ell m n}\in\mathbb{R},
\label{eq:def-delta}
\end{equation}
where $(\boldsymbol{\theta}_n)$ denotes the parameters that generate the fractional frequency shift. We focus on $p\gg1$ so that $\delta_{\ell m n}^{\mathrm{th}}$ is given by 
\begin{equation}
\delta^{\mathrm{th}}_{\ell m n}(\boldsymbol{\theta})
\;=\;
-\frac{1}{2\pi\, f^{\mathrm{GR}}_{\ell m n}\, \tau^{\mathrm{GR}}_{\ell m n}\, \sqrt{2p-3}}\;,
\label{eq:model-delta}
\end{equation}
Here $f^{\mathrm{GR}}_{\ell m n}$ and $\tau^{\mathrm{GR}}_{\ell m n}$ denote the GR-predicted mode frequency and damping time, respectively, and they are functions of the final mass and spin, $(M_f,\chi_f)$.  We use the QNM samples derived from the LVK remnant posterior based on the IMR quasi-circular, spin-precessing \texttt{NRSur7dq4} model~\cite{Varma:2019csw}.  Ideally the baseline QNM quantities would be computed from the same \texttt{pSEOBNR} analysis used for the deviation parameters; the reduced products used here do not provide that joint information.  The agreement of the remnant predictions from the two analyses indicates that the resulting difference is negligible at the precision relevant here~\cite{KAGRA:2025oiz}.

For each mode $i\in\{220,440\}$, we build a one-dimensional probability density $p_{\mathrm{obs},i}(\delta)$ by applying a kernel density estimator to the LVK posterior samples of the dimensionless frequency shift $\delta_i$.  We then marginalize once over the common remnant posterior,
\begin{equation}
\mathcal{L}(p)
~\propto~
\int \! \mathrm{d}M_f\,\mathrm{d}\chi_f\;
p_{\mathrm{GR}}(M_f,\chi_f\mid d)\,
\prod_{i\in\{220,440\}}
p_{\mathrm{obs},i}\!\left[
\delta^{\mathrm{th}}_i\!\left(p;M_f,\chi_f\right)
\right].
\label{eq:like-int}
\end{equation}
Eq.~(\ref{eq:like-int}) is not the most general likelihood since the product of one-dimensional deviation densities neglects correlations of each $\delta f_i$ with the remaining waveform parameters, as well as residual correlations between the deviation parameters.  This approximation is operationally required by the reduced one-dimensional products used in our analysis.  It is also physically motivated for a GW250114-like signal, because the high-SNR inspiral and merger tightly constrain the GR binary and remnant parameters before the post-merger deviations are measured. For instance, Ref.~\cite{Ghosh:2021mrv} explicitly finds negligible correlations between the non-GR QNM-frequency parameter and the binary parameters when appreciable pre-merger information is present (GW150914), while such correlations become important for a signal lacking a measurable pre-merger part (GW190521).  The GW250114 \texttt{pSEOBNR} analysis likewise reports only minimal correlations among the mode-deviation parameters~\cite{LIGOScientific:2025obp}.  Although a future reanalysis using the full joint samples would be preferable, we expect the neglected correlations to be subdominant here.
We adopt a prior uniform in $\log_{10}p$ over the interval $[1,10]$. The posterior is
\begin{equation}
p\!\left(p\,\middle|\,\mathrm{data}\right)
~\propto~
\mathcal{L}(p)\,\pi(p),
\label{eq:posterior-final}
\end{equation}
where $\pi(p)$ denotes the prior.
 
The resulting marginalized posterior for $\log_{10}p$ is illustrated in Fig.~\ref{fig:SMB_two_panel}. These constraints quantify the level to which the GW250114 ringdown allows deviations by the SMB effect. A comparison with Fisher-matrix forecasts for next-generation detectors is provided and the details are discussed in Sec.~\ref{sec:forecast}. It can be seen from Fig.~\ref{fig:SMB_two_panel} that the LVK posterior (blue and orange) rises toward large $p$ and effectively excludes small values, yielding an approximate lower bound
$\log_{10}p \gtrsim 2$. This indicates that the small frequency shift due to the GW250114 requires the suppression due to SMB to be very small. Moreover, we find that the posterior of $p$ obtained by combining the 220 and 440 modes is nearly identical to the one only using the 220 mode, indicating that the constraint is dominated by the 220 mode. This is as expected because, for GW250114, the 220 mode is especially well constrained, unlike the 440 mode.

\section{Fisher Matrix Forecasts for GW250114-like Events}\label{sec:forecast}
To estimate the expected precision with which future ground-based detectors such as CE and ET could measure potential deviations from the GR QNM frequencies, we performed a Fisher-matrix forecast \cite{Cutler:1994ys} based on a GW250114-like event.

We adopted the \texttt{IMRPhenomD} waveform model \cite{Husa:2015iqa,Khan:2015jqa,lalsuite,swiglal} as implemented in the 
\texttt{GWFish} package \cite{Dupletsa:2022scg}, and modified it to include a phenomenological frequency shift
$\delta f$ in the ringdown mode (the 220 mode). Specifically, the ringdown frequency in the model, originally given by $f_{\mathrm{RD}}$, was replaced by
$f_{\mathrm{RD}} (1 + \delta f)$
where $\delta f$ is a dimensionless fractional deviation from the GR-predicted value. 

For the fiducial binary parameters, we adopted the median value of  GW250114 as reported by the LVK:
a chirp mass of $\mathcal{M} = 28.9\,M_\odot$, mass ratio $q = 0.96$, 
component spins $a_1 = 0.24$, $a_2 = 0.26$, and a luminosity distance corresponding 
to redshift $z = 0.09$ (approximately $400\,\mathrm{Mpc}$) \cite{KAGRA:2025oiz,LIGOScientific:2025obp}. In addition to these standard parameters, we introduced $\delta f$ as a new parameter in the waveform generation, while keeping the rest of the \texttt{IMRPhenomD} model unchanged. The null forecasts shown in Fig.~\ref{fig:SMB_two_panel} use a GR fiducial waveform, $\delta f_{\mathrm{inj}}=0$, and the Fisher matrix is evaluated at that GR point.

We computed the Fisher information matrix using the \texttt{GWFish} framework \cite{Dupletsa:2022scg}:
\begin{equation}
    \Gamma_{ij} = 
    \left( \frac{\partial h}{\partial \theta_i} \middle| 
           \frac{\partial h}{\partial \theta_j} \right),
\end{equation}
where $(a|b)$ denotes the standard noise-weighted inner product over frequency, and $\theta_i$ are the model parameters.
The signal-to-noise ratio (SNR) and Fisher matrix were computed for CE/ET detector configurations.
The corresponding inverse Fisher matrix provides the covariance matrix
$\Sigma = \Gamma^{-1}$, from which we derived the $1\sigma$ uncertainties.

To visualize the expected parameter correlations, we sampled $5\times10^5$
points from the multivariate Gaussian defined by $\Sigma$ and the fiducial mean values. 
The right panel of Fig.~\ref{fig:SMB_two_panel} shows the marginalized distribution of $\delta f$.

\begin{figure*}[ht]
    \centering
    \includegraphics[width=0.88\textwidth]{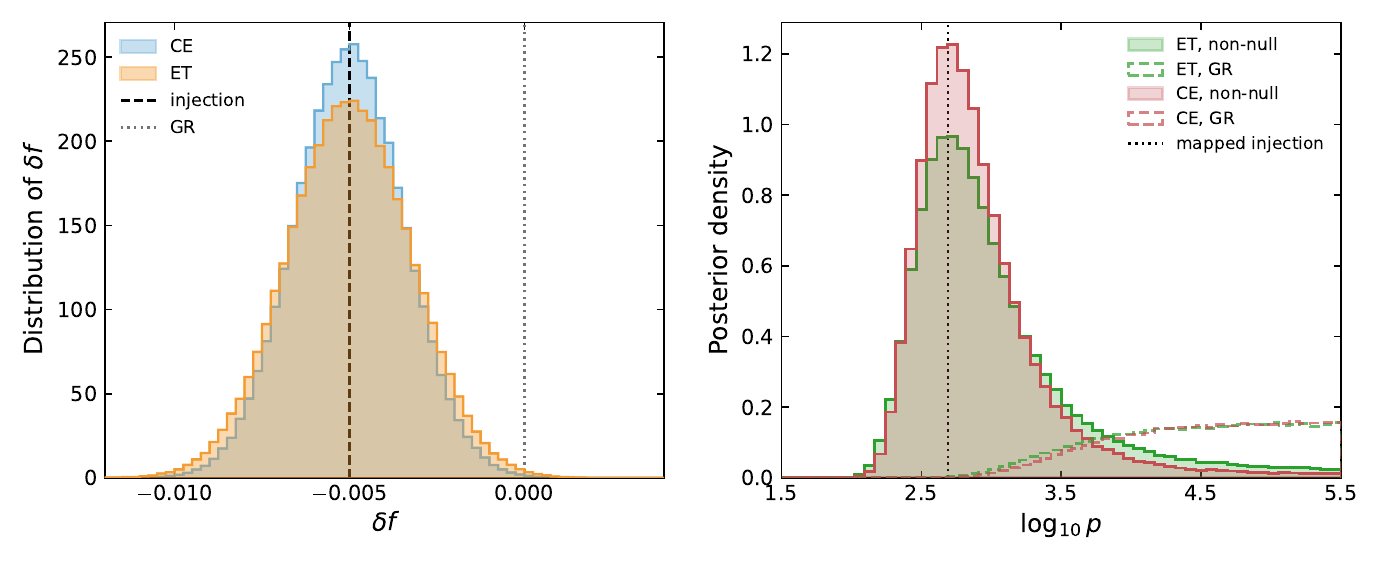}
    \caption{\label{fig:nonnull_injection}
Non-null Fisher analysis.  \textbf{Left panel:} Marginalized ET and CE histograms obtained by evaluating the Fisher matrix at the injected shift $\delta f_{\mathrm{inj}}=-0.005$; the black dashed and gray dotted lines mark the non-null and GR values, respectively.  \textbf{Right panel:} Corresponding posterior histograms for $\log_{10}p$, compared with the GR-injection posteriors.  The black dotted line marks the value $\log_{10}p_{\mathrm{inj}}=2.690$ obtained by mapping the $-0.5\%$ shift through Eq.~(\ref{eq:model-delta}) using the GR value of 220-mode.}
\end{figure*}
The analysis indicates that, for a GW250114-like event observed by CE/ET, 
the deviation parameter $\delta f$ can be constrained to a few thousandth, depending on the assumed network sensitivity. This result highlights the capability of next-generation detectors to perform precision tests of the deviation of the QNM frequencies with unprecedented accuracy.

Again, we perform another Bayesian inference based on the Fisher-matrix forecast result, as shown in the right panel of Fig.~\ref{fig:SMB_two_panel} for $\delta f$ and also the corresponding bound of $p$ in the left panel.
It can be seen that the CE/ET Fisher forecast (red/green) results of $p$ move to higher values, indicating a prospective bound
$\log_{10}p \gtrsim 3$, one order of magnitude better than the LVK result. Because the SMB suppression factor scales as $[(1-f/f_R)\mu]^{1/(p-1)}$, these bounds force the power-law index $1/(p-1)$ to be very small. Since the observed frequency shift of GW250114-like events is small, it implies $\mathcal{O}(1)$ suppression near $f_R$.

As a non-null validation, we repeat the complete Fisher calculation with $\delta f_{\mathrm{inj}}=-0.005$.  We find the shift corresponds to $p_{\mathrm{inj}}\simeq4.90\times10^2$, or $\log_{10}p_{\mathrm{inj}}=2.690$.  At the non-null point we obtain $\sigma_{\delta f}=1.77\times10^{-3}$ for ET and $1.56\times10^{-3}$ for CE, so the injected shift differs from zero by $2.82\sigma$ and $3.21\sigma$, respectively.  As shown in Fig.~\ref{fig:nonnull_injection}, both $p$ posteriors peak at $\log_{10}p\simeq2.69$, although with a noticeable large-$p$ tail, whereas the GR-injection posteriors rise toward the $p\rightarrow\infty$ limit.  Thus the analysis recovers a finite injected value and exhibits behavior distinct from the null case.

\section{Summary and discussion}\label{sec:sum}
We have proposed and tested a minimal, data-driven framework to probe the SMB with the ringdown signal of binary BH merger. In the SMB picture, assisted gaplessness makes a large family of memory modes nearly degenerate when a merger drives the system. The information load resists depletion into radiation, pushing the effective resonance toward the infrared and narrowing the resonance window. We encoded this physics in a single-mode ringdown model in which the spectral peak is slightly redshifted while the overall intensity near the unburdened resonance is suppressed. Using LVK posteriors for GW250114 and combining the (220) and (440) modes, our Bayesian analysis yields a lower bound $\log_{10}p \gtrsim 2$. On the other hand, Fisher-matrix forecasts for a CE/ET observation of a GW250114-like source indicate prospective constraints around $\log_{10}p \gtrsim 3$. Together, these results show that current data already disfavor strongly burdened responses with rapidly reopening gaps, and that next-generation sensitivity will further constrain the parameters.

Our phenomenological burdened ringdown model implements the SMB-induced suppression by multiplying the single-mode ringdown spectrum with the suppression factor,
$S(f;\mu,p)$, and then infers a peak redshift $\delta f$.
However, we need to remind the readers that the suppression factor given by Eq.~(\ref{Sfactor}) is an upper bound~\cite{Dvali:2025sog}. Treating the upper bound as an equality yields the least suppression. Consequently, our lower bound on $p$ is conservative. If the true suppression is stronger than Eq.~(\ref{Sfactor}), the same data would imply a larger lower bound on $p$.

Our Bayesian analysis only considered the QNM frequency shifts. This approach discards information contained in a full waveform analysis. This restriction must not be interpreted as a prediction that $\delta\tau=0$ for the SMB effect: the SMB suppression bound predicted in~\cite{Dvali:2025sog} gives a non-Lorentzian ringdown spectral profile, but which we can infer a frequency shift, but the current model does not allow us to obtain a clearly defined shift to the QNM damping times.  Deriving a definite relation $\delta\tau^{\mathrm{th}}(p)$, requires additional microscopic or time-domain input beyond the present phenomenological model. On the other hand, the peak shift of the burdened spectrum is independent of the information-load parameter $\mu$ where $\mu$ rescales the overall intensity. Consequently, frequency-shift data alone can only bound $p$, instead of $(\mu,p)$ jointly.
A full waveform analysis, or a measurement of the ringdown amplitudes, is needed to lift this degeneracy and obtain simultaneous constraints on both $\mu$ and $p$. A fully rigorous approach would require a complete analysis of the waveform of GW250114, comparing an unburdened ringdown-only waveform with a ringdown waveform modified by the SMB effect in order to infer constraints on both model parameters $\mu$ and $p$.

Our analysis models the ringdown as a linear superposition of Lorentzian modes. In reality, the late-time signal can exhibit nonlinear behavior 
\cite{Okuzumi:2008ej,Mitman:2022qdl,Guerreiro:2023gdy,Cheung:2023vki,Zhu:2024rej,Redondo-Yuste:2023seq,Khera:2023oyf,Cardoso:2024jme,He:2025ydh}. Given that current observations show that the $(220)$ mode dominates the ringdown SNR while subdominant modes such as $(440)$ are measured with larger uncertainties (e.g., \cite{KAGRA:2025oiz,LIGOScientific:2025obp}), our lower bound on $p$ is primarily driven by the $(220)$ mode frequency shift and should be relatively robust to modest nonlinear corrections.

The analysis presented here is motivated by the SMB proposal \cite{Dvali:2025sog}, which is currently not yet part of a broadly established theoretical framework. Accordingly, the results of this work should be interpreted as phenomenological and model-dependent constraints within a minimal SMB framework, rather than as definitive bounds. Our goal is to provide a brief demonstration of how BH spectroscopy can be used to confront such proposals with GW data, and to shed some light on which regions of parameter space are testable with current and next-generation detectors under explicit assumptions.

In summary, our constraint on $p$ should be interpreted as a conservative lower bound derived from a peak-only analysis. A full SMB waveform analysis could be of interest and is expected to extract joint $(\mu,p)$ constraints.

\acknowledgments
We thank Gia Dvali for insightful comments on the swift memory burden effect.
C.Y. acknowledge the financial support provided under the European Union’s H2020 ERC Advanced Grant “Black holes: gravitational engines of discovery” grant agreement no. Gravitas–101052587. Views and opinions expressed are however those of the author only and do not necessarily reflect those of the European Union or the European Research Council. Neither the European Union nor the granting authority can be held responsible for them. We acknowledge support from the Villum Investigator program supported by the VILLUM Foundation (grant no. VIL37766) and the DNRF Chair program (grant no. DNRF162) by the Danish National Research Foundation.
R.B. acknowledges financial support provided by FCT – Fundação para a Ciência e a Tecnologia, I.P., through the ERC-Portugal program Project ``GravNewFields''. We also thank the Fundação para a Ciência e Tecnologia (FCT), Portugal, for the financial support to the Center for Astrophysics and Gravitation (CENTRA/IST/ULisboa) through grant No. UID/PRR/00099/2025 (\nolinkurl{https://doi.org/10.54499/UID/PRR/00099/2025}) and grant No. UID/00099/2025 (\nolinkurl{https://doi.org/10.54499/UID/00099/2025}).
%
%\end{acknowledgements}

%%%%%%%%%%%%%%%%%%%%%
\bibliographystyle{JHEP}
\bibliography{ref}
%%%%%%%%%%%%%%%%%%%%%

\end{document}